\documentclass[prb,twocolumn,showpacs,floatfix]{revtex4}
\usepackage{graphicx,color}
\usepackage{amssymb}
\usepackage{amsmath}
\usepackage{xspace}
\usepackage{url}

\begin{document} 
\title{Competing chiral and multipolar electric phases in the extended Falicov-Kimball model}
%
\author{B. Zenker and H. Fehske}
\affiliation{Institut f{\"ur} Physik,
             Ernst-Moritz-Arndt-Universit{\"a}t Greifswald,
             D-17489 Greifswald, Germany}
\author{C. D. Batista}
\affiliation{Theoretical Division, Los Alamos National Laboratory,
             Los Alamos, New Mexico 87545, USA}

\date{\today}
\begin{abstract}
We study the effects of interband hybridization within the framework of
an extended Falicov-Kimball model with itinerant $c$ and $f$ electrons.
An explicit interband hybridization  breaks the U(1) symmetry associated with the conservation of the 
difference between the total number of particles in each band. 
As a result, the degeneracy between multipolar electric 
and chiral orderings is lifted. 
We analyze the weak- and strong-coupling limits of the $c$-$f$ electron 
Coulomb interaction at zero temperature, 
and derive the corresponding mean-field quantum phase diagrams at half-filling for a model defined on a square lattice.
\end{abstract}
\maketitle

\section{Introduction}  

The Falicov-Kimball model~\cite{FK69,RFK70} 
(FKM) was primarily introduced to describe the metal-insulator transition 
of the mixed-valence compound  SmB$_6$. Later on, the model became widely 
accepted as a minimal Hamiltonian for studying several strongly correlated 
electron systems,~\cite{FZ03,GM96,P77,NA85,FNM01,FD01} in particular, 
heavy fermion compounds.~\cite{MCA80,GB01,L01}
In its original form, the FKM 
contains  an itinerant $c$ band of electrons that interact via a local Coulomb 
repulsion with localized $f$ electrons. 
The spin degree of freedom of the electrons is not included. The
local $f$ electron number is strictly conserved
and $c$-$f$ electron coherence cannot be established.~\cite{SB88}
An explicit hybridization between $f$ and $c$ orbitals provides 
an opportunity to overcome this shortcoming.~\cite{KMM76,POS96b} 
More recently, it was shown that a finite $f$ electron bandwidth 
also induces $c$-$f$ electron 
coherence, i.e., it can lead to an excitonic condensate
even in absence of an explicit interband hybridization.~\cite{Ba02b,BGBL04} 

These extended versions of the FKM were used to
substantiate the exciting idea of electronic 
ferroelectricity (EFE).~\cite{POS96b,Ba02b,BGBL04,Fa99,Fa08}
The ferroelectric phase only appears when the $c$ and $f$ orbitals 
have opposite parity under spatial inversion. 
The concomitant spontaneous breaking of inversion symmetry  
results from a nonvanishing 
 average of $\langle c^\dagger f^{}\rangle$. Since 
this expectation value 
corresponds to (excitonic) pairing of electrons and holes from different bands, 
the appearance of EFE is directly related 
with the formation of an excitonic insulator (EI).~\cite{Mo61,Kn63,HR68}

The FKM with two dispersive bands, the so-called 
extended Falicov-Kimball model (EFKM), was 
studied previously for describing different properties 
of the EI phase.~\cite{SC08,IPBBF08,Br08,ZIBF10,PBF10}
However, as it was shown for the case of opposite-parity 
orbitals,~\cite{Ba02b,BGBL04} the inclusion of a
finite interband hybridization can be very relevant because it removes 
the U(1) symmetry associated with the conservation of the 
difference between the total number of 
particles in each band: $N_c-N_f$. In particular, this hybridization 
term is certainly relevant when the ground-state
of the EFKM corresponds to an  excitonic condensate. For the EFKM with 
interband hybridization (HEFKM), the  excitonic condensate
is in general replaced by Ising-type phases that only break 
discrete symmetries of the Hamiltonian. 

In this paper we present a mean-field study of the influence of an explicit hybridization
on the symmetry-broken states that can take place for the HEFKM.
To determine the ground-state quantum phase diagram of the HEFKM
in the strong- and weak-coupling limits of  $c$-$f$ electron interaction, 
we assume that the interband hybridization amplitudes 
are small compared to the intraband hopping (transfer) integrals. 
Given the nature of the discrete symmetries of the HEFKM, 
the natural  ground-state candidates 
are chiral phases (CHPs)  and states with multipolar electric orderings. 

\section{Model}
By expressing the orbital flavor as a pseudospin variable, 
$c^{\dagger}_{i} \equiv c^{\dagger}_{{i} \uparrow}$ and 
$f^{\dagger}_{i} \equiv c^{\dagger}_{{i} \downarrow}$ ,
the Hamiltonian takes the form
\begin{eqnarray}
{\cal H} &=& \sum_{i,\sigma} \varepsilon_{\sigma} c_{i \sigma}^\dagger c_{i \sigma}^{\;} 
+ \sum_{\langle i  j \rangle, \sigma} t_{\sigma} \left( c_{i \sigma}^\dagger c_{j \sigma}^{\,}
+{\rm H.c.} \right)
\nonumber \\
&+&U \sum_{i} n_{{i}\uparrow} n_{{i}\downarrow}
+v_0 \sum_{{i}} \left(c_{{i}\uparrow}^\dagger c_{{i}\downarrow}^{\,} +{\rm H.c.} \right)
\nonumber  \\
&+&  v_{\uparrow\downarrow} \sum_{\langle i j\rangle}
\left(c_{{i}\uparrow}^\dagger c_{{j} \downarrow}^{\,}+{\rm H.c.} \right)
\nonumber \\
&+& v_{\downarrow\uparrow} \sum_{\langle i j\rangle} 
\left(c_{{i}\downarrow}^\dagger c_{{j} \uparrow}^{\,}+{\rm H.c.} \right)\,. 
\label{HEFKM}
\end{eqnarray}
Here $\langle i  j \rangle$ indicates that $i$ and $j$ are
 nearest-neighbor sites. The fermionic operators $c_{{i}\sigma}^{(\dagger)}$ 
annihilate (create) an electron on the spin $\sigma$ Wannier orbital of the lattice site ${\bf R}_{i}$.
The lattice has a total number of $N$ sites, and 
$n_{{j}\sigma}=c_{{j}\sigma}^\dagger c_{{j}\sigma}^{}$
is the particle number operator for site $j$
($\sigma=\{\uparrow,\downarrow\} $). 
$\varepsilon_\sigma$ denotes the on-site energy for each orbital, 
$t_{\sigma}$ are the intraband  hopping amplitudes, 
$U$ is the local interorbital Coulomb interaction strength, and
$v_\gamma$ are the interband hybridization amplitudes, where $\gamma=0$ for 
on-site hybridization and $\gamma=\{\uparrow\downarrow,\downarrow\uparrow\}$
for intersite hybridization.  
The EFKM is recovered from Eq.~\eqref{HEFKM} by setting  
$v_\gamma=0$. In this limit, 
the model has a continuous U(1) symmetry, which is removed by the 
inclusion of an explicit hybridization. The discrete symmetries that remain
for the more general HEFKM  are spatial inversion and 
time-reversal invariance. 

The pseudospin language of Eq.~\eqref{HEFKM} unveils the similarity of ${\cal H}$ with other generic many-body Hamiltonians.  The EFKM ($v_\gamma=0$) becomes an asymmetric Hubbard model, i.e., a single band model for  electrons  with a spin-dependent dispersion.  We will still use the name "Falicov-Kimball model" to indicate that the pseudospin degree of freedom represents a physical orbital degree of freedom.
From now on, we will consider  that ${\cal H}$  is defined on a
square lattice and $\langle n_{i\uparrow} + n_{i\downarrow} \rangle =1 $ (half-filled band case). We will also restrict to zero temperature and measure 
all energies in units of $t_{\uparrow}=1$.
Finally, we will assume that the Wannier functions
of $c$ and $f$ orbitals, $\phi_{\uparrow}({\bf r}-{\bf R}_i)$ and $\phi_{\downarrow}({\bf r}-{\bf R}_i)$, are real.

\section{Order parameters}

In the rest of the paper we will refer to the pseudospin simply as ``spin.'' 
The spin representation used in Eq.~\eqref{HEFKM} unveils the SU(2) structure of this internal degree of freedom.
This degree of freedom is the only one that survives at low energies in the large $U/|t_{\sigma}|$ limit. Consequently,
the three different local or real-space  order parameters  correspond to the three components of the local spin variable,
\begin{equation}
{\boldsymbol {\cal S}}_{j} = \frac{1}{2} \sum_{\sigma,\sigma'} c^{\dagger}_{j \sigma} {\boldsymbol \sigma}_{\sigma \sigma'} c^{}_{j \sigma'}, 
\label{spin}
\end{equation}  
where ${\boldsymbol \sigma}$ is the vector of the Pauli matrices. More complicated (or higher order) real-space order parameters involve products of spin operators in
more than one unit cell.

A real space modulation of $\langle {\cal S}^{z}_j \rangle$  leads to {\it orbital ordering}. Here we will only consider the
ordering wave vector  ${\bf Q}=(\pi,\pi)$ that leads to  {\it staggered orbital ordering} (SOO) 
because the effective interaction is antiferromagnetic between nearest-neighbors  and the lattice under consideration is bipartite.  
The corresponding order parameter is
\begin{equation}
 \delta_{\text{SOO}}  = \sum_{j} e^{i{\bf Q} \cdot {\bf R}_j}\langle {{\cal S}}_j^z \rangle \;. 
\label{SOO}
\end{equation}

If the two orbitals have opposite parity, a nonzero $\langle {\cal S}_{j}^x\rangle$ implies the presence of a spontaneous local electric polarization
that turns out to be uniform  for the HEFKM. This is the  EFE that was found in previous works for particular limits
of the HEFKM.~\cite{POS96b,Ba02b,BGBL04} The uniform electric polarization is given by 
\begin{equation}
\langle{\boldsymbol{ \cal P}}\rangle = {\bf p} \sum_{j}  \langle {{\cal S}}_{j}^x \rangle
\label{Pol}
\end{equation}
with the interband dipole matrix element
\begin{equation}
{\bf p} = 2e\int  d^3r \;\phi_\uparrow ({\bf r}) \; {\bf r} 
\;\phi_\downarrow ({\bf r}) \;,\label{dipole_me} 
\end{equation}
where $e$ is the electron charge.  This phase breaks the spatial inversion symmetry of ${\cal H}$.

If the two orbitals have {\it the same parity} (for instance $s$ and $d$ orbitals), a nonzero modulation of $\langle {\cal S}^{x}_{j} \rangle$ 
corresponds to an {\it electric quadrupole density wave} (EQDW) as long as the tensor
\begin{equation}
q^{\nu \nu'}_{\sigma \sigma'} = e \int d^3r\, \phi_{\sigma}({\bf r})  r_{\nu}  r_{\nu'}    \phi_{\sigma'}({\bf r}) 
\label{quad}
\end{equation} 
is nonzero for ${\sigma'={\bar \sigma}} \equiv -\sigma$  ($\nu,\nu'=\{x,y,z\}$).
In case the tensor ${\bf q}_{\sigma {\bar \sigma}}$ [Eq.~\eqref{quad}] vanishes, 
one has to look for the lowest order electric  multipole that has a nonzero matrix element
between the orbitals $\phi_{\uparrow}({\bf r})$ and $\phi_{\downarrow}({\bf r})$. In the rest of this paper, we will assume 
that the tensor ${\bf q}_{\sigma {\bar \sigma}}$ does not vanish. In second quantization, the local electric quadrupole 
tensor on the unit cell $j$ is given by the operator,
\begin{equation} 
{\boldsymbol {\cal Q}}_{j} = {\bf q}_{\uparrow\uparrow} n_{j \uparrow} + {\bf q}_{\downarrow\downarrow}
n_{{j}\downarrow} + 2 {\bf q}_{\uparrow \downarrow} S^x_{j}.
\label{quadop}
\end{equation} 
We note that ${\bf q}_{\uparrow \downarrow}={\bf q}_{\downarrow \uparrow}$.
The corresponding quadrupolar order parameter in momentum space is given by
\begin{equation} 
\langle {\boldsymbol {\cal Q}}_{\bf k} \rangle = \sum_{j} e^{i {\bf k} \cdot {\bf R}_j}  \langle {\boldsymbol{\cal Q}}_{j} \rangle \;.
\label{quadopmom}
\end{equation} 
Again, for the Hamiltonian under consideration, the wave vector of the electric quadrupolar ordering is ${\bf k}={\bf Q}$. 
Equation~\eqref{quadop} implies that a nonzero modulation of the $x$-spin component, $\langle {\cal S}^{x}_j \rangle$, 
corresponds to an EQDW. While translational symmetry is broken in this phase, time-reversal and spatial inversion symmetries are conserved. 
We note that the first two terms of Eq.~\eqref{quadop} imply that orbital ordering will also lead to an  EQDW. However, as it is also clear from
Eq.~\eqref{quadop}, the quadrupolar tensors associated with the ordering along the $z$ and $x$ axes are different. The quadrupolar electric moment
that is modulated under staggered orbital ordering (staggered $z$ component) corresponds to a linear combination of the tensors 
${\bf q}_{\uparrow \uparrow}$ and ${\bf q}_{\downarrow \downarrow}$. On the other hand, the staggered ordering of the $x$-spin 
component involves a modulation of a quadrupolar electric tensor proportional to ${\bf q}_{\uparrow \downarrow}$ (hybridization-induced
quadrupolar electric moment). In order to simplify 
the notation, we will use ``EQDW'' to denote the staggered ordering of the $x$-spin component and SOO for the 
the staggered ordering of the $z$-spin component.

Finally, a nonzero $\langle {\cal S}^{y}_j \rangle$ implies the spontaneous
emergence of a current-density distribution between the two orbitals of the unit cell $j$.~\cite{HR68,B81} This can be easily
verified if we neglect the overlap between orbitals that belong to different unit cells. In that case, the current-density
operator at a point ${\bf r}$ near  ${\bf R}_j$ is given by
\begin{equation}
{\bf j}_{j} ({\bf r}) = \frac{\hbar}{m_e} S^{y}_{j} \sum_{\sigma}  \sigma  \phi_{{\bar \sigma}}({\bf r} - {\bf R}_j) 
{\boldsymbol \nabla}_{\bf r} \phi_{\sigma}({\bf r } - {\bf R}_j),
\label{curr} 
\end{equation}
where $m_e$ is the electron mass and the prefactor $\sigma$ takes the value $+1$ for $\uparrow$ and 
$-1$ for $\downarrow$. We note that this current density flows between the two orbitals of the same unit cell 
(these are atomic currents when the two orbitals belong to the same ion), in contrast to the 
orbital currents found in Ref.~\onlinecite{BBMK08} that flow between different unit cells.
This CHP breaks time-reversal symmetry
and the physical order parameter is the lowest order nonzero multipole of the current-density distribution given by  Eq.~\eqref{curr}. 
The chiral ordering is {\it staggered}  for orbitals with the same parity and {\it uniform} for orbitals with opposite parity.
For instance, if we are considering two $p$ orbitals, the staggered chiral ordering of
the $y$-spin component corresponds to  orbital antiferromagnetism, because the current distribution given by
Eq.~\eqref{curr} generates a net magnetic dipole moment.  Since we will consider the general case of
any arbitrary pair of orbitals, we will use ``CHP'' to denote the uniform ordering of the $y$-spin component (same parity orbitals)
and ``staggered chiral phase'' (SCHP) to denote the staggered ordering.

\section {Strong-coupling regime}

For the case $|t_{\sigma}|,|v_\gamma| \ll U$, we can perform a large-$U$ 
expansion, thereby reducing the HEFKM [Eq.~(\ref{HEFKM})] to an effective 
strong-coupling Hamiltonian, $ {\cal H}^{\rm sc}$, that reproduces the low-energy spectrum of the original model. 

A large on-site Coulomb interaction splits the spectrum of  
the HEFKM Hamiltonian  into 
high- and low-energy parts. For $t_{\sigma}=v_\gamma=0$,  the 
lowest-energy subspace is generated by the $2^N$ states 
that have one electron per site. 
The high-energy subspaces are separated by energy gaps equal to
$Un_d$, where $n_d$ is the number of double occupied sites. 
For nonzero $t_{\sigma}$ and $v_\gamma$, the electrons are no longer
completely localized at their ions, i.e., an electron can  
gain kinetic energy by visiting virtually a 
neighboring site.  Since we are considering the half-filled band case
(one particle per site), the low-energy effective model 
becomes a spin Hamiltonian ${\cal H}^{\rm sc}$. The expression for ${\cal H}^{\rm sc}$ up to second order in the
kinetic-energy terms is
\begin{align}
{\cal H}^{\rm sc} =& \sum_{\langle i  j \rangle} ( 
J_{xx} {{\cal S}}^x_{i} {{\cal S}}^x_{j} 
+J_{yy} {{\cal S}}^y_{i} {{\cal S}}^y_{j}
+J_{zz} {{\cal S}}^z_{i} {{\cal S}}^z_{j})
\nonumber \\
+&  \sum_{\langle i j\rangle} (J_{xz} {{\cal S}}^x_{i} {{\cal S}}^z_{j} +  
J_{zx} {{\cal S}}^z_{i } {{\cal S}}^x_{j} + C)\nonumber \\
+& \sum_{i} ( B   {{\cal S}}^z_{i} + 2 v_0  {{\cal S}}^x_{i})\,,
\label{H_Spin}
\end{align}	
where 
\begin{align}
J_{xx} =& \frac{4}{U} \left( t_{\uparrow} t_{\downarrow} + v_{\uparrow \downarrow} v_{\downarrow \uparrow} \right) \label{Jxx1} \;,\\
J_{yy} =& \frac{4}{U} \left( t_{\uparrow} t_{\downarrow} - v_{\uparrow \downarrow} v_{\downarrow \uparrow} \right) \label{Jyy1} \;,\\
J_{zz} =& \frac{2}{U} \left( t_{\uparrow}^2 +t_{\downarrow}^2 - v_{\uparrow \downarrow}^2 - v_{\downarrow \uparrow}^2 \right) 
\label{Jzz1} \;,\\
J_{xz} =& \frac{4}{U} \left( t_{\uparrow} v_{\downarrow \uparrow} - t_{\downarrow} v_{\uparrow \downarrow} \right) \label{Jxz1} \;,\\
J_{zx} =& \frac{4}{U} \left( t_{\uparrow} v_{\uparrow \downarrow} - t_{\downarrow} v_{\downarrow \uparrow} \right) \label{Jzx1} \;,\\
C =& \varepsilon_{\uparrow}+\varepsilon_{\downarrow} - \frac{1}{2U} \left( t_{\uparrow}^2 
+ t_{\downarrow}^2 + v_{\uparrow \downarrow}^2 
+ v_{\downarrow \uparrow}^2 \right) \label{c1} \;,\\
B =& \varepsilon_\uparrow - \varepsilon_\downarrow \label{az1} \;.
\end{align}
 It is well known that the half-filled isotropic Hubbard model can be mapped 
on an effective Heisenberg model in the limit of a large Coulomb repulsion. 
For the more general EFKM, the intraband hopping amplitudes and the 
different on-site potentials lead to an effective XXZ model 
in a magnetic field $B$ along the $z$  axis. $B$ is simply the 
energy difference between the two orbitals.~\cite{Ba02b} As expected, 
the interband hybridization  of the HEFKM 
generates anisotropic terms that explicitly 
break the U(1) invariance under uniform  spin rotations
about the $z$ axis. 
While the intersite hybridization leads to anisotropic exchange terms, 
the on-site hybridization leads to a Zeeman coupling to a uniform field 
along the $x$ axis.
 
For low enough values of $B$ and no interband hybridization, 
the ground-state of ${{\cal H}^{\rm sc}}$ exhibits SOO.
The simple reason is that $J_{zz} \geq |J_{xx,yy}|$, i.e., the 
effective XXZ model is easy-axis.~\cite{Ba02b} If $J_{zz}$ is 
significantly larger than $|J_{xx,yy}|$, the SOO remains robust 
when the interband hybridization is included. 
Clearly, there exists a critical value of $B$ that leads 
to a spin-flop transition to an ordered phase in the $XY$ plane with a uniform
component along the $z$ axis (canted $XY$ phase). 
In absence of interband hybridization, the U(1) invariance of  the 
EFKM implies that  spin component perpendicular to the applied field  can point 
along any direction of the $XY$ plane. In other words, there is 
a continuous ground-state degeneracy that includes the 
EQDW (EFE) and SCHP (CHP) for orbitals with the same (opposite) parity. 
In this case, the inclusion of interband hybridization is very 
relevant because it lifts the continuous degeneracy and stabilizes 
only one of the two possible Ising-type orderings
(along the $x$ or $y$ spin direction). 

For orbitals with opposite parity, we have $t_{\uparrow} t_{\downarrow}<0$, $ v_{\downarrow \uparrow}=-v_{\uparrow \downarrow}$ and $v_0=0$. 
These relationships are derived from simple symmetry considerations.
In this case $J_{xx} = -\frac{4}{U} ( |t_{\uparrow} t_{\downarrow}| + v_{\uparrow \downarrow}^2)$, $J_{yy} = -\frac{4}{U} (| t_{\uparrow} t_{\downarrow}| - v_{\uparrow \downarrow}^2)$, and $J_{zx}=-J_{xz}$. Since $J_{xx}, J_{yy}<0$ and  $|J_{xx}|>|J_{yy}|$, the energy is minimized by a ferromagnetic alignment of the spins along the $x$ direction that corresponds to an EFE phase. Since this was previously shown in Ref.~\onlinecite{Ba02b}, from now on we will concentrate on the case of
equal parity orbitals. In this case, we have $t_{\uparrow} t_{\downarrow}>0$, $ v_{\downarrow \uparrow}=v_{\uparrow \downarrow}$, and $v_0$ can be nonzero if the two orbitals belong
to different ions. Then $J_{xx} > J_{yy}$  and $J_{zx}=J_{xz}$. Since $J_{xx}>J_{yy}>0$, $v_{\uparrow\downarrow}$ favors 
{\it staggered} Ising-type ordering along the $x$  direction, while 
$v_0$ favors a {\it uniform} polarization along the $x$ direction and, consequently, a {\it staggered} 
Ising-type ordering along the $y$  direction. 
In other words, the interband hybridization can stabilize an EQDW or a SCHP depending on the ratio between the on-site and intersite hybridization amplitudes.

We introduce now the mean-field variational states and the corresponding energies 
for the three order parameters that we introduced in the previous section.

(i) {\it SOO}. This phase has a staggered 
spin component along the $z$-direction, and a uniform component along the $x$ direction 
that can be induced by the on-site hybridization term $v_0$,
\begin{align}
\langle{ \boldsymbol{\cal S}}_{j}\rangle_{\theta_1} &= S\left(\sin{\theta_1},   0, \cos{\theta_1} e^{i{\bf Q} {\bf R}_j} \right) \label{S_SOO}\;, \\
E_0^{\rm SOO} =& -DNS^2J_{zz}\cos^2\theta_1+DNS^2J_{xx}\sin^2\theta_1 \nonumber \\
+&2v_0SN\sin\theta_1+DNC\;.
\label{E_SOO}
\end{align}

(ii) {\it SCHP}. This phase has a staggered 
 spin component along the $y$ direction and uniform polarizations along the $x$ and $z$ directions,
\begin{align}
\langle{ \boldsymbol{\cal S}}_{j}\rangle_{\theta_1,\theta_2} &= S\left(\sin\theta_1 \cos\theta_2,  e^{i{\bf Q} {\bf R}_j} \sin\theta_1 \sin\theta_2, \cos\theta_1 \right) \label{S_chiral}\;, \\
E_0^{{\rm SCHP}} =&\; DNS^2 \left( (J_{xx}+J_{yy})\cos^2\theta_2(1-\cos^2\theta_1)- J_{yy} \nonumber \right.\\
+& \left. (J_{yy}+J_{zz}) \cos^2\theta_1  + 2J_{xz}\sin\theta_1 \cos\theta_1 \cos\theta_2 \right) \nonumber \\
+& DNC + NS B \cos\theta_1 + 2NSv_0 \sin\theta_1 \cos\theta_2 \label{E_chiral} \;.
\end{align}

(iii) {\it EQDW}. In this case the staggered spin component is aligned along
the $x$ direction and there is a uniform component along the $z$ direction induced by $B$ (the $y$ component vanishes),
\begin{align}
\langle{\boldsymbol{\cal S}}_{j} \rangle_{\theta_1} &= S\left( e^{i {\bf Q} {\bf R}_j} \sin\theta_1, 0 , \cos\theta_1 \right) \label{S_staggeredx} \;,\\
E_0^{\rm EQDW} =&\; DNC + N B S \cos\theta_1 - DNJ_{xx} S^2\sin^2 \theta_1 \nonumber \\
+& DN J_{zz} S^2\cos^2 \theta_1  \;.
\label{E_staggeredx}
\end{align}

In all cases we have $D=2$, ${\bf Q}=(\pi,\pi)$, and $S=1/2$. By minimizing the respective energies 
with respect to $\theta_1$ and $\theta_2$, we determine the quantum 
phase diagram as a function of the band-structure parameters.

\begin{figure}[h]
\centering
\includegraphics[width=\linewidth]{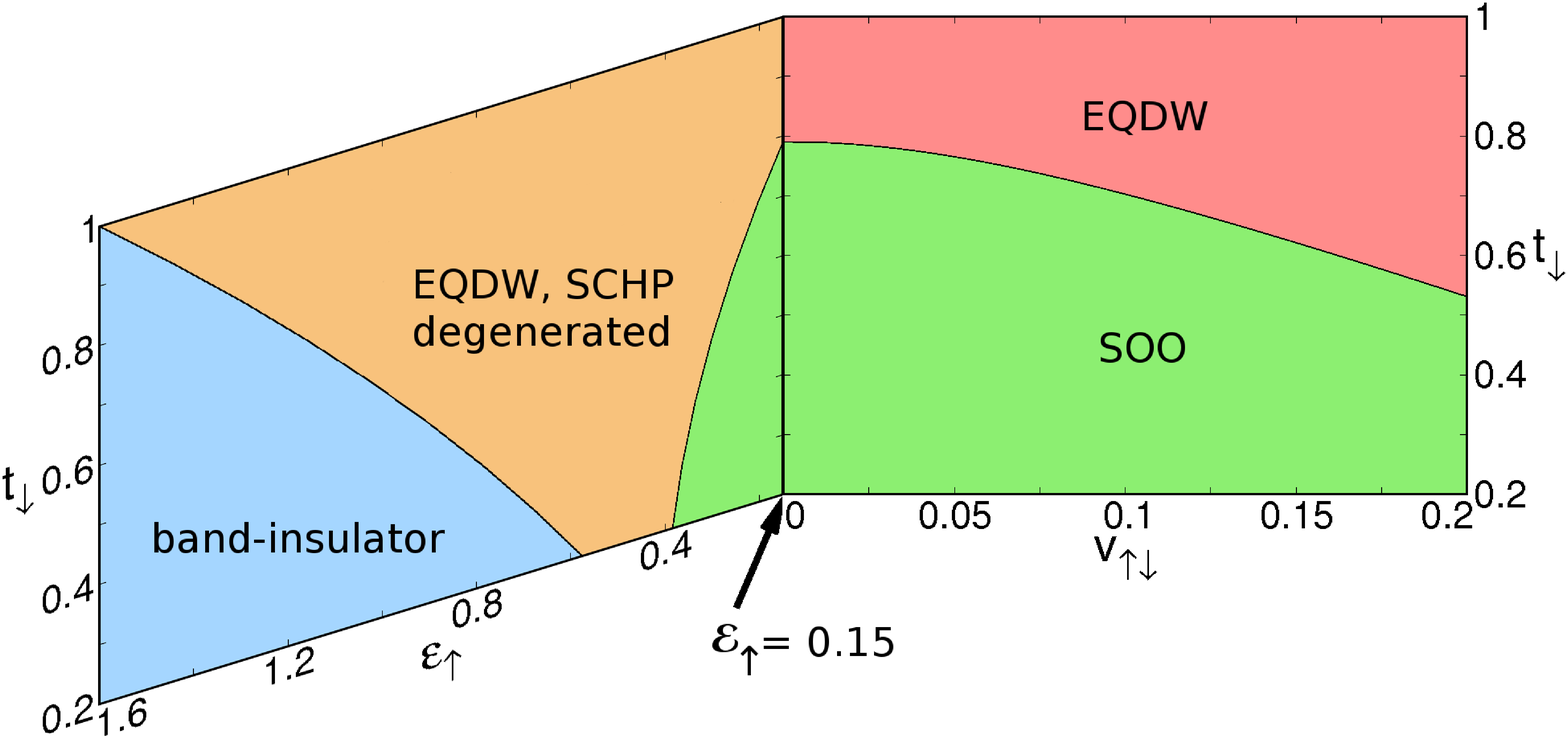}
{\flushleft (a)\\}
\includegraphics[width=\linewidth]{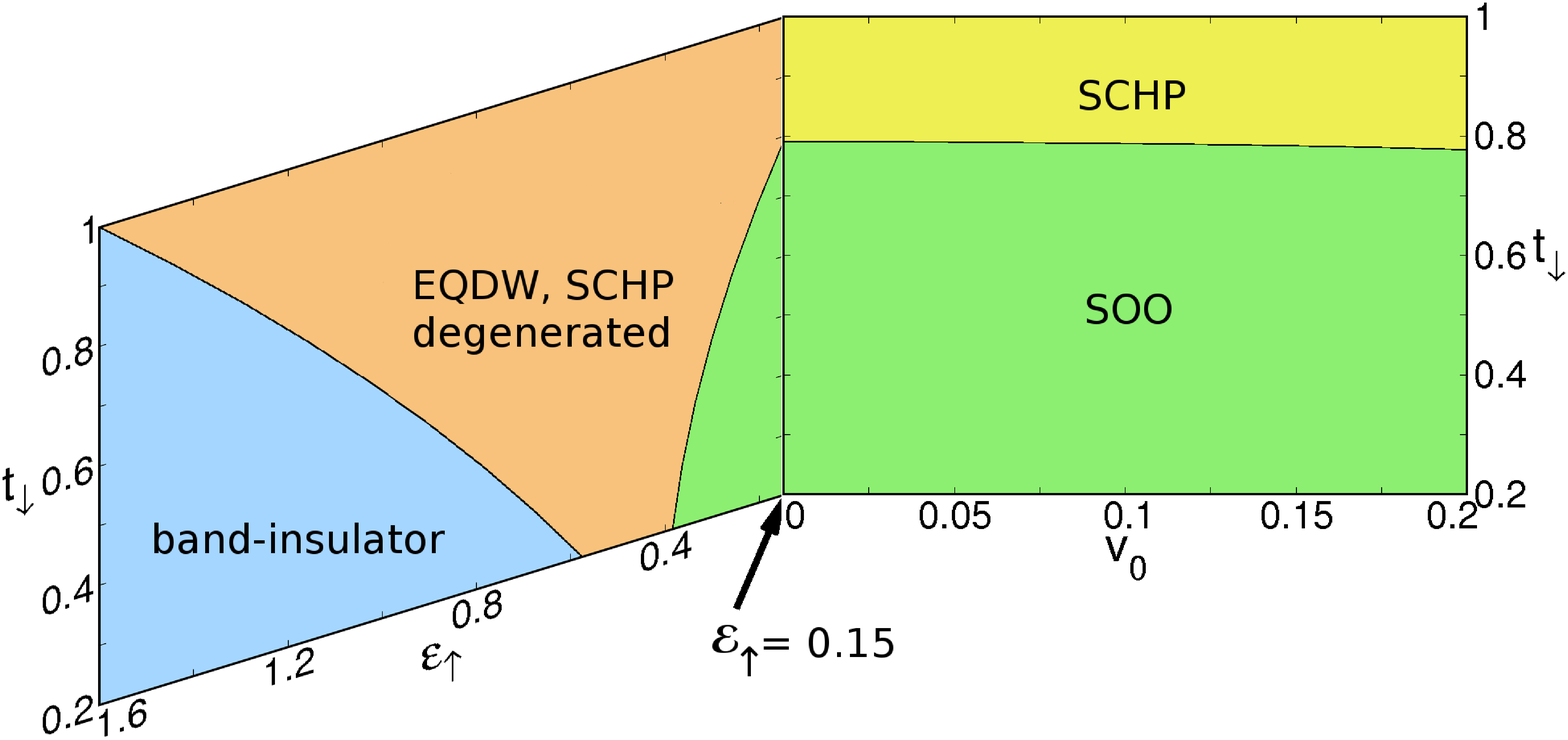}
{\flushleft (b)\\}
\includegraphics[width=0.55\linewidth]{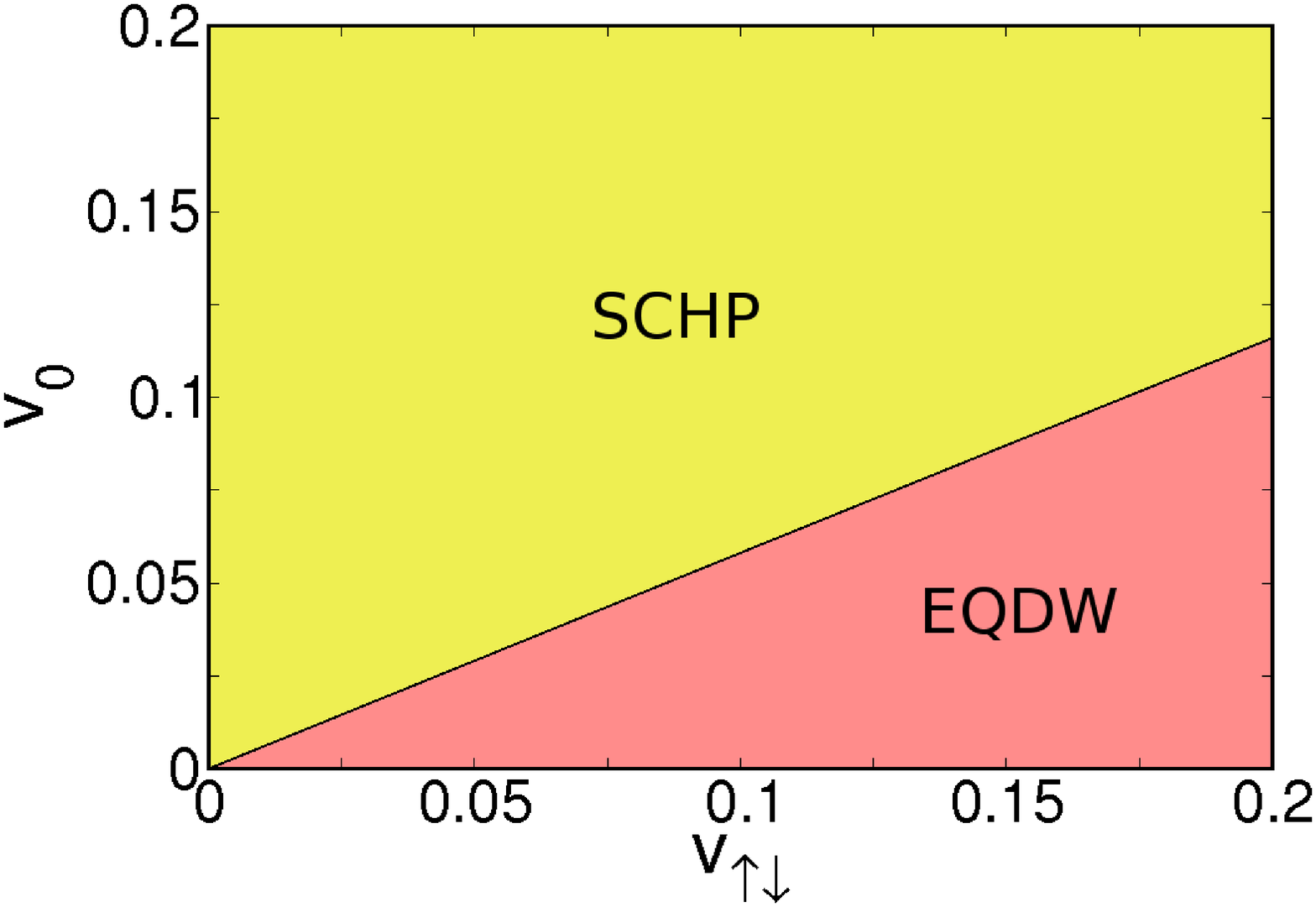}
{\flushleft (c)\\}
\caption{(Color online)  
Ground-state phase diagram of the 2D EFKM in the strong-coupling regime. 
Band-structure parameters are $\varepsilon_\downarrow =0.0$, 
$t_{\uparrow}=1.0$, and  $U=10$. Left-hand side diagrams [in panels 
(a) and (b)] 
give results for the nonhybridized EFKM ($v_0=0$ and 
$v_{\uparrow\downarrow}=0$), while right-hand side diagrams show the dependence 
on the (a) intersite hybridization $v_{\uparrow\downarrow}$ and 
 (b) on-site hybridization $v_0$ for $\varepsilon_\uparrow=0.15$.
Panel (c) gives the stability region of the staggered chiral phase 
and the electric quadrupole density wave in dependence on $v_0$ 
and $v_{\uparrow\downarrow}$ for $\varepsilon_\uparrow=0.5$, 
$t_{\downarrow}=0.5$.}
\label{fig_largeU}
\end{figure}

 Figure~\ref{fig_largeU}(a) shows the effect of a 
finite intersite hybridization amplitude, $ v_{\uparrow\downarrow}$, on the
quantum phase diagram of the EFKM. The intersite hybridization 
stabilizes the EQDW relative to the SCHP. On the other hand, the EQDW 
is also favored relative to the SOO because 
$ v_{\uparrow\downarrow}$ decreases the value of $J_{zz}$
and simultaneously increases the value of $J_{xx}$ [see Eqs.~\eqref{Jxx1} and \eqref{Jzz1}]. Figure~\ref{fig_largeU}(b) illustrates 
the effect of a finite on-site hybridization. 
In this case, the SCHP is favored relative 
to the EQDW. In contrast to $v_{\uparrow\downarrow}$, $v_0$ 
does not change  the transition point between the SCHP and the 
SOO. The simple reason is that $v_0$ does not affect the exchange constants. It just generates an effective pseudomagnetic field along the $x$ 
axis that leads to a finite
canting angle in both phases.
Figure~\ref{fig_largeU}(c) shows the phase diagram as a function 
of $v_0$ and $v_{\uparrow\downarrow}$ for a large enough value 
of $B=0.5$ and same parity orbitals. Again, we can see that a finite on-site hybridization $v_0$ strengthens the SCHP, while the intersite hybridization $v_{\uparrow\downarrow}$ stabilizes the EQDW. Within our simple mean-field approximation the boundary between these two phases is a straight line.

\section{Weak-coupling regime}
It has been shown in Ref.~\onlinecite{Fa08} that the mean-field ground-state phase diagram of the 2D EFKM agrees almost perfectly with the one obtained by a constrained path Monte Carlo technique, even in the intermediate coupling regime. This agreement motivates us to perform a Hartree-Fock decoupling of the HEFKM to explore the quantum phase diagram for small $U/|t_{\sigma}|$.

The weak-coupling analysis requires to express the relevant order parameters in momentum space. 
In particular, the Fourier components  of  $\langle {{\cal S}}_{j}^x\rangle$ and $\langle {{\cal S}}_{j}^y\rangle$ can be represented as a complex number,
\begin{eqnarray}
\Delta_{\bf Q} &=& |\Delta_{\bf Q}|e^{i\varphi}=
\frac{U}{N}\sum_{\bf k} \langle c_{{\bf k}+{\bf Q}\uparrow}^\dagger c_{{\bf k}\downarrow}^{} \rangle 
\label{EI_OP} \\
&=& \frac {U}{N} \sum_{j} e^{i{\bf QR}_j} \left( \langle{{\cal S}}_{j}^x\rangle + i \langle {{\cal S}}_{j}^y \rangle \right) \;
\label{E_I_OP_real} \\
&=& \frac {U}{\sqrt{N}} (\langle {\cal S}^x_{\bf Q} \rangle + i \langle {\cal S}^y_{\bf Q} \rangle ),
\label{EI_OP_S}
\end{eqnarray}
where
\begin{align}
c_{{\bf k}\sigma}^{\dagger} =& \frac{1}{\sqrt{N}} \sum_{j} e^{i {\bf k} \cdot {\bf R}_j} c_{j \sigma}^{\dagger}, 
 \\
{\cal S}^{\nu}_{{\bf k}} =& \frac{1}{\sqrt{N}} \sum_{j} e^{i {\bf k} \cdot {\bf R}_j} {\cal S}^{\nu}_{j}. 
\end{align}
The ordering wave vector ${\bf Q}$ determines the modulation of the real-space order parameter. According to 
our strong-coupling analysis, we have ${\bf Q}=(\pi,\pi)$ for orbitals with the same parity and ${\bf Q}=(0,0)$
for orbitals with opposite parity. Again, the U(1) invariance of the EFKM ($v_\gamma=0$) implies that the energy does not depend on $\varphi$. 
Consequently, there is an infinite number of ground-states with $\Delta_{\bf Q} \neq 0$ that results from the spontaneous  
U(1) symmetry breaking of the EFKM (excitonic condensate). A finite interband hybridization ($v_\gamma\neq0$) removes the continuous U(1)
symmetry and lifts the $\varphi$ degeneracy of the HEFKM ground-state. 
 
For  orbitals with opposite parity, the hybridization in momentum space takes the form $v_{\bf k} = 2i v_{\uparrow \downarrow} (\sin k_x + \sin k_y)$ 
and the EFE state has a lower energy than the CHP. This is in agreement with the result for the FKM extended by a (small) intersite hybridization in Ref.~\onlinecite{POS96b}. Therefore, from now on we will focus only on the equal parity case. 
The Hartree-Fock decoupling suggested  by Eq.~\eqref{EI_OP} gives
\begin{align}
{\cal H}^{\rm wc} =& \sum_{{\bf k},\sigma} \bar{\varepsilon}_{{\bf k}\sigma}\, c_{{\bf k}\sigma}^\dagger c_{{\bf k}\sigma}^{} 
+ \sum_{{\bf k},\sigma} v_{{\bf k}}\, c_{{\bf k}\sigma}^\dagger c_{{\bf k}\,-\sigma}^{}\nonumber \\
-&\sum_{{\bf k}} \Delta_{\bf Q}\, c_{{\bf k}\downarrow}^\dagger c_{{\bf k}+{\bf Q}\uparrow}^{} 
- \sum_{{\bf k}} \Delta_{\bf Q}^\ast\,c_{{\bf k}+{\bf Q}\uparrow}^{\dagger} c_{{\bf k}\downarrow}^{} 
\label{HEFKM_HF}
\end{align}
with
\begin{align}
\bar{\varepsilon}_{{\bf k}\sigma} =& \varepsilon_\sigma + U n_{-\sigma} + 2t_{\sigma} (\cos k_x +\cos k_y) \label{ek} \;,\\
v_{{\bf k}} =& v_0+2v_{\uparrow\downarrow}  (\cos k_x +\cos k_y) \label{Vk} \;,\\
n_\sigma =& \frac{1}{N}\sum_{\bf k} \langle c_{{\bf k}\sigma}^{\dagger} c_{{\bf k}\sigma}^{}\rangle \label{n_k} \;,\\
\Delta_{\bf Q} =& \frac{U}{N} \sum_{{\bf k}} \langle c_{{\bf k}+{\bf Q}\uparrow}^\dagger c_{{\bf k}\downarrow}^{} \rangle \label{Delta} \;,\\
\Delta_{\bf Q}^\ast =& \frac{U}{N} \sum_{{\bf k}} \langle c_{{\bf k}\downarrow}^\dagger c_{{\bf k}+{\bf Q}\uparrow}^{} \rangle \label{Delta*} \;.
\end{align}
The mean-field Hamiltonian~(\ref{HEFKM_HF}) can be easily 
diagonalized by the canonical transformation~\cite{BZGB05}
\begin{align}
{\cal C}_{{\bf k},m} = u_{{\bf k},m} c_{{\bf k}\uparrow}^{} + v_{{\bf k},m} c_{{\bf k}\downarrow}^{} + \tilde{u}_{{\bf k},m} c_{{\bf k}+{\bf Q}\uparrow} ^{}
+ \tilde{v}_{{\bf k},m} c_{{\bf k}+{\bf Q}\downarrow}^{} \label{canonTransf} \;,
\end{align}
where $m=1,2,3,4$. 
The coefficients are solutions of the
associated Bogoliubov  de~Gennes equations,
$
{\cal H}_{\bf k}^{\rm wc} \Psi_{{\bf k},m} = E_{{\bf k},m} \Psi_{{\bf k},m}\;, 
$
with
\begin{align}
{\cal H}_{\bf k}^{\rm wc}=&\left(\begin{matrix} \bar{\varepsilon}_{{\bf k}\uparrow} & v_{{\bf k}} & 0 & -\Delta_{\bf Q}^\ast \\ v_{{\bf k}} & \bar{\varepsilon}_{{\bf k}\downarrow} & -\Delta_{\bf Q} & 0 \\
 0 & -\Delta_{\bf Q}^\ast & \bar{\varepsilon}_{{\bf k}+{\bf Q}\uparrow} & v_{{\bf k}+{\bf Q}} \\
  -\Delta_{\bf Q} & 0 & v_{{\bf k}+{\bf Q}} & \bar{\varepsilon}_{{\bf k}+{\bf Q}\downarrow} \end{matrix}\right) 
 \label{HEFKM_HF_matrix}
\end{align}
and $\Psi_{{\bf k},m}=(u_{{\bf k},m} , v_{{\bf k},m} , \tilde{u}_{{\bf k},m}, \tilde{v}_{{\bf k},m})^T$.
The energy per site results as
\begin{align}
\frac{E_0^{\rm wc}}{N} =& \frac{1}{N}\sum_{{\bf k},m}{'}  \, E_{{\bf k},m} f(E_{{\bf k},m})- U n_\uparrow n_\downarrow + \frac{1}{U}|\Delta_{\bf Q}|^2 
\label{energy_EI} \;,
\end{align}
where  $f(E_{{\bf k},m})$ is the Fermi function containing the new quasiparticle energies $E_{{\bf k},m}$ and the prime denotes 
that the ${\bf k}$ summation extends over 
the magnetic Brillouin zone only. 
The chemical potential $\mu$ is determined 
by the condition 
\begin{equation}
1=\frac{1}{N}\sum_{{\bf k},m}{'} f(E_{{\bf k},m})\,. \label{chemPot}
\end{equation}

Next we consider the mean-field decoupling that leads to SOO. 
In this case,  we introduce the possibility of a periodic modulation in the electronic density with independent amplitudes 
for each spin polarization,
\begin{align}
\langle n_{{i}\sigma}\rangle =& n_\sigma + \delta_\sigma \cos({\bf QR}_i), \label{density} 
\end{align}
with
\begin{align}
\delta_\sigma =& \frac{1}{N} \sum_{{\bf k}}{} \langle c_{{\bf k}\sigma}^\dagger c_{{\bf k}+{\bf Q}\sigma}^{} \rangle \label{CDW_param}\;.
\end{align}
The associated Bogoliubov de~Gennes equations are
$
{\cal H}_{\bf k}^{\rm SOO} {\Psi}_{{\bf k},m} = {E}_{{\bf k},m}^{\rm SOO} \Psi_{{\bf k},m},
$
with
\begin{align}
{\cal H}_{\bf k}^{\rm SOO}=&\left(\begin{matrix} \bar{\varepsilon}_{{\bf k}\uparrow} & v_{\bf k} & U\delta_\uparrow &  0 \\
v_{\bf k}  & \bar{\varepsilon}_{{\bf k}\downarrow} & 0 & U\delta_\downarrow \\
U\delta_\uparrow & 0 & \bar{\varepsilon}_{{\bf k} +{\bf Q}\uparrow} & v_{{\bf k}+{\bf Q}} \\
 0 & U\delta_\downarrow & v_{{\bf k}+{\bf Q}} & \bar{\varepsilon}_{{\bf k}+{\bf Q}\downarrow} \end{matrix}\right) \;.
 \label{HEFKM_HFSOO_matrix}
\end{align}
The SOO order parameter becomes
\begin{equation}
\delta_{\text{SOO}} = \frac{\delta_{\uparrow} - \delta_{\downarrow}}{2}. \label{SOO_OP_HF}
\end{equation} 
For asymmetric bands, $t_{\uparrow} \neq t_{\downarrow}$, the presence 
of a nonzero SOO leads to  a secondary charge-density-wave (CDW) 
order, whose order parameter is given by
\begin{equation}
\delta_{\text{CDW}}  = \frac{\delta_{\uparrow} + \delta_{\downarrow}}{2}. \label{CDW_OP_HF}
\end{equation} 
This secondary CDW provides a simple way of detecting the SOO in real materials. 
The mean-field energy per site that results from such a kind of  SOO is
\begin{align}
\frac{E_0^{\rm SOO}}{N} =& \frac{1}{N}\sum_{{\bf k},m}{'}{E}_{{\bf k},m}^{\rm SOO} f(E_{{\bf k},m}^{\rm SOO}) - U n_\uparrow n_\downarrow - U \delta_\downarrow \delta_\uparrow \label{E_SOO_HF} \;.
\end{align}

By solving the self-consistency Eqs.~\eqref{Delta} and~\eqref{CDW_param}, and comparing the corresponding  mean-field energies given by Eqs.~\eqref{energy_EI} and \eqref{E_SOO_HF}, we compute the ground-state phase diagram.  

\begin{figure}[h]
\centering
\includegraphics[width=\linewidth]{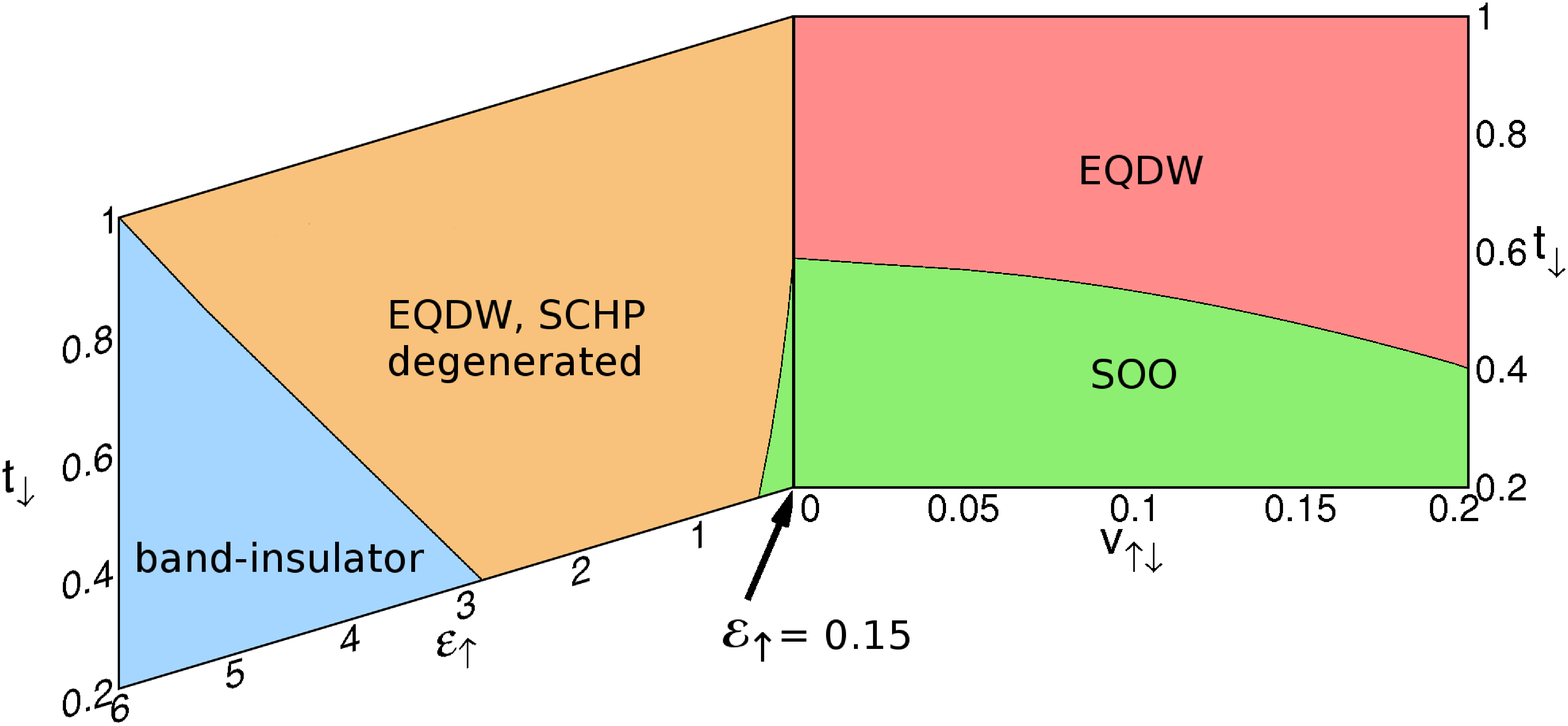}
{\flushleft (a)\\}
\includegraphics[width=\linewidth]{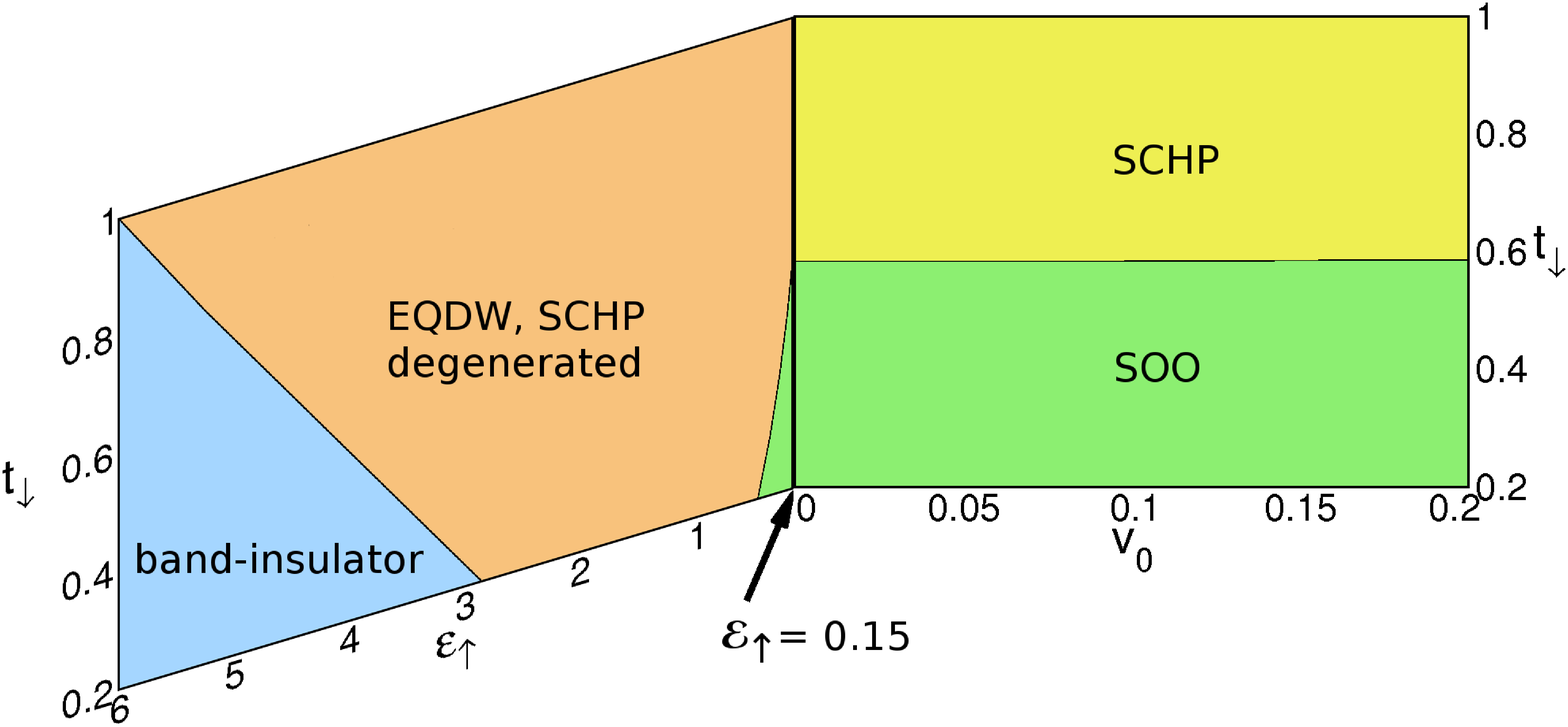}
{\flushleft (b)\\}
\includegraphics[width=0.55\linewidth]{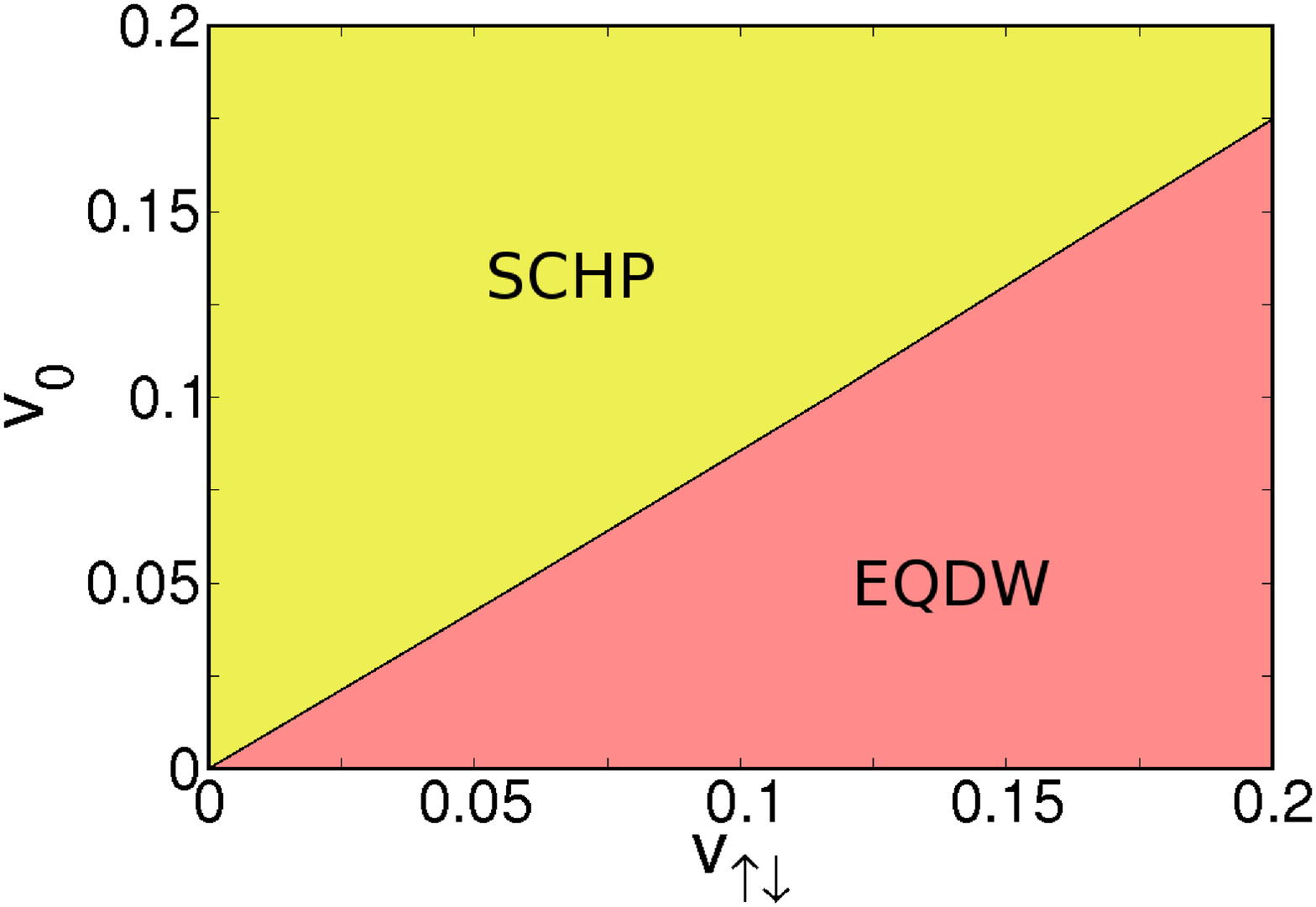}
{\flushleft (c) \\}
\caption{(Color online) 
Ground-state phase diagram of the 2D EFKM in the weak-coupling regime. 
Band-structure parameters are $\varepsilon_\downarrow =0.0$, 
$t_{\uparrow}=1.0$, and  $U=2$. Left-hand side diagrams [in panels~(a) and (b)] 
give results for the nonhybridized EFKM ($v_0=0$ and 
$v_{\uparrow\downarrow}=0$), while right-hand side diagrams show the dependence 
on the (a) intersite hybridization $v_{\uparrow\downarrow}$ and 
 (b) on-site hybridization $v_0$ for $\varepsilon_\uparrow=0.15$.
Panel (c) gives the stability region of the staggered chiral phase 
and the electric quadrupole density wave in dependence on $v_0$ 
and $v_{\uparrow\downarrow}$ for $\varepsilon_\uparrow=0.5$, 
$t_{\downarrow}=0.5$.}
\label{fig_HF}
\end{figure}

Figure~\ref{fig_HF} is the weak-coupling counterpart of 
Figure~\ref{fig_largeU}. Like for the strong-coupling regime, 
Fig.~\ref{fig_HF}(a) shows that an increasing value of $v_{\uparrow\downarrow}$ 
narrows the SOO phase while the region of the EQDW phase is enlarged. 
On the other hand, the on-site hybridization $v_0$ favors the SCHP relative to the  EQDW and it does not have a noticeable effect on the 
transition line between the SOO and the SCHP [see Fig.~\ref{fig_HF}(b)]. 
This also coincides with the strong-coupling results. 
Figure~\ref{fig_HF}(c)  shows the stability regions of the EQDW and SCHP as a function of the hybridization amplitudes for a large enough $|\varepsilon_\uparrow-\varepsilon_\downarrow|=0.5$. {Qualitatively, the result of our Hartree-Fock approach is similar to the one obtained from the
strong-coupling analysis [see Fig.~\ref{fig_largeU}(c)]. However, a more quantitative analysis shows that the area of stability for the  SCHP is reduced 
relative to the strong-coupling result.} A large Coulomb repulsion inhibits hopping processes and consequently reduces the influence of 
the intersite hybridization $v_{\uparrow\downarrow}$ relative to the  effect of the on-site hybridization $v_0$.

\section{Conclusions}
The Hamiltonian that we considered in this work is a very simple extension of the Falicov-Kimball model. 
In spite of its simplicity, we have shown that this model leads to a very rich quantum phase diagram that contains all the 
possible local order parameters (three different components of the local spin ${\boldsymbol{\cal S}_j}$)   considered in Sec.~III. 
The ordering wave vector ${\bf Q}$ is selected by the nesting property of the noninteracting Fermi surface in the weak-coupling limit and by the antiferromagnetic nature of the exchange interactions on a bipartite lattice in the strong-coupling limit. Most notably, 
the stability of the different broken symmetry states is very sensitive to a few band-structure parameters. According to these results, 
it is necessary to have very accurate information about the band-structure properties near the Fermi energy to predict the correct ordered state. 
In particular, if the two orbitals have different angular momentum (like $s$ and $d$ orbitals), the SCHP may remain hidden to most of the experimental probes. The simple reason is that the spontaneous current-density distribution given by Eq.~\eqref{curr} has no net magnetic moment. Consequently, this phase can only be detected by using an experimental probe that couples to the lowest nonzero multipole of the current-density distribution. 
The SCHP becomes stable
above a critical value of the on-site hybridization as long as the diagonal energy difference between the two orbitals $|\varepsilon_c-\varepsilon_f|$
is also larger than a minimal value. 

Although all the calculations of this work were done for $D=2$,  we do not expect any qualitative change for  $D > 2$.
The obtained consistency between the weak- and the strong-coupling approaches suggests that our results are robust.
In particular, the absence of geometric frustration in the strong-coupling regime, whose weak-coupling counterpart is the nesting
property of the Fermi surface, facilitates the search for the broken symmetry state that minimizes the energy for each set
of Hamiltonian parameters. For orbitals with opposite parity under spatial inversion, we confirmed that the ferroelectric phase has always a lower 
energy than the chiral phase. For orbitals with the same parity, we found that the stabilization of the  electric quadrupole density wave or the staggered chiral phase depends strongly on the dominant
interband hybridization. The on-site hybridization, that is only allowed when the two orbitals belong to different ions, favors the
staggered chiral phase, while a nearest-neighbor interband hybridization favors the electric quadrupole density wave.  

\section{Acknowledgments}
This work was supported by the Deutsche Forschungsgemeinschaft 
through  SFB 652, B5, and by the NNSA
of the U.S. DOE at the Los Alamos National Laboratory under Contract No. DE-AC52-06NA25396.
B.Z. and H.F. are grateful for the hospitality provided at the Los Alamos National Laboratory.


\end{document}